\documentclass[reprint,preprintnumbers,nofootinbib,superscriptaddress,amsmath,amssymb,aps,prd]{revtex4-2}
\pdfoutput=1
\usepackage{graphicx,hyperref,braket,orcidlink, algorithm2e, amssymb, amsmath, dsfont, subfigure, lineno, bm, slashed, multirow, slashed, booktabs, subfigure}
\usepackage{tikz}
\usepackage{quantikz}
\usetikzlibrary{positioning,decorations.pathreplacing,decorations.markings,calc,arrows.meta}
\usepackage[nolist,nohyperlinks]{acronym}
\graphicspath{{figs/}}
\hypersetup{
	bookmarks=true,
	unicode=true,
	pdftoolbar=true,
	pdfmenubar=true,
	pdffitwindow=false,
	pdfstartview={FitH},
	pdftitle={Quantum simulation of real-time Yukawa dynamics with qubits and qumodes},
	pdfauthor={J.Y.~Araz, J.~Forshaw, R.~Jenkinson}, 
	pdfsubject={Quantum simulation of real-time Yukawa dynamics with qubits and qumodes},
	pdfcreator={J.Y.~Araz, J.~Forshaw, R.~Jenkinson},
	pdfproducer={}, 
	pdfkeywords={},
	pdfnewwindow=true,
	colorlinks=true,
	linkcolor=blue,
	citecolor=olive,
	filecolor=magenta,
	urlcolor=cyan
}

\definecolor{org}{HTML}{FF4F00}

\newcommand{\Hf}{H_{f}}
\newcommand{\Hs}{H_{s}}
\newcommand{\Hint}{H_{\rm int}}
\newcommand{\sx}{\sigma^{x}}
\newcommand{\sy}{\sigma^{y}}
\newcommand{\sz}{\sigma^{z}}

\definecolor{cgreen}{HTML}{2CA02C}
\definecolor{cblue}{HTML}{1F77B4}
\definecolor{corange}{HTML}{FF7F0E}
\definecolor{cyellow}{HTML}{FFD700}
\definecolor{cgrey}{HTML}{7F7F7F}
\definecolor{cred}{HTML}{D62728}
\definecolor{cpurple}{HTML}{9467BD}

\begin{document}

% Find a better title
%\title{Towards simulating scattering in hybrid qubit--qumode quantum hardware for lattice field theory}
% \title{Scalar decay and resonant scattering beyond one loop for qubit--qumode quantum simulators}
\title{Quantum simulation of real-time Yukawa dynamics with qubits and qumodes}

\author{Jack Y. Araz\orcidlink{0000-0001-8721-8042}}
\email{j.araz@ucl.ac.uk}
\affiliation{Department of Physics and Astronomy, University College London, London, WC1E 6BT, UK}
\affiliation{Department of Engineering, City St George’s, University of London, London, EC1V 0HB, UK}

\author{Jeffrey Forshaw\orcidlink{0000-0001-9590-7390}}
\email{jeffrey.forshaw@manchester.ac.uk}
\affiliation{Department of Physics and Astronomy, University of Manchester, Manchester, UK}

\author{Ross Jenkinson\orcidlink{0000-0003-3199-6704}}
\email{ross.jenkinson@manchester.ac.uk}
\affiliation{Department of Physics and Astronomy, University of Manchester, Manchester, UK}

\begin{abstract}
Real-time quantum simulation of field theories offers access to scattering and decay dynamics, circumventing the limitations of Euclidean lattice methods. However, representing bosonic fields on qubits requires a finite-dimensional approximation and can entail costly multi-qubit operations. Hybrid qubit--qumode architectures avoid this by encoding bosonic fields directly in oscillators. We formulate scattering in the $(1+1)$-dimensional Yukawa theory for such a device, with staggered fermions on qubits and the scalar on qumodes. Trapped ions are a natural candidate, with the fermions on the internal states of the ions and the scalar on their collective motion, so that the Yukawa coupling becomes a native spin-dependent displacement. Emulating the device with tensor networks, we study the decay of a single scalar particle and fermion--antifermion scattering around the resonance region, using observables that can be measured on the device natively. We compare the results with those obtained using lattice perturbation theory. The emulation is able to resolve corrections beyond the Dyson summation of the one-loop self-energy.
\end{abstract}

\maketitle

\tableofcontents

%%%%%%%%%%%%%%%%%%%%%%%%%%%%%%%%%%%%%%%
\section{Introduction}
\label{sec:intro}
%%%%%%%%%%%%%%%%%%%%%%%%%%%%%%%%%%%%%%%

Scattering in quantum field theory is a real-time process, yet Monte Carlo sampling of the real-time path integral is hindered by the sign problem~\cite{Bauer:2022hpo,Banuls:2019bmf}. Quantum computers offer an alternative by implementing Hamiltonian time evolution~\cite{Lloyd1996,JordanLeePreskill2014}, and methods have been developed to prepare fermionic and hadronic wave packets~\cite{Chai:2023qpq,Davoudi:2024wyv}. Scattering has been demonstrated in scalar and Ising field theories using more than a hundred qubits~\cite{Zemlevskiy:2024vxt,Farrell:2025nkx}, and scattering in the interacting Thirring model has been simulated on 40 qubits~\cite{Chai:2025kbi}. Meson collisions have been studied on superconducting and trapped-ion hardware~\cite{Schuhmacher:2025ehh,Davoudi:2025rdv}. In one spatial dimension, tensor-network methods provide classical benchmarks for particle production and entanglement generation~\cite{Rigobello:2021fxw,Belyansky:2023rgh}. Resonances have also been studied directly in the time domain using matrix product states (MPS)~\cite{Jha2025} and through a trapped-ion simulation of a resonant toy model~\cite{Balaji:2026zqb}.

Qubits are the standard choice for these simulations, and each fermionic mode maps exactly onto one qubit. A bosonic field, however, can only be represented approximately on a finite qubit register, through Fock-space truncation or field digitisation~\cite{Somma2015, Macridin2018PRL, Macridin2018PRA, KlcoSavage2019}. Other approaches discretise or reduce the gauge-field basis~\cite{Hackett2019,Ji2020, DavoudiRaychowdhuryShaw2020, Ciavarella2021}, or retain states below an energy cutoff~\cite{Rychkov:2014eea, Hogervorst:2014rta, Ingoldby:2024fcy, Ingoldby:2025bdb}. These approaches can be effective at low energies, but maintaining accuracy at higher energies or over longer evolution times can require more resources~\cite{Tong2022, SahinogluSomma2021, Hanada2023}. Binary encodings can also require multi-qubit arithmetic circuits to implement boson dynamics~\cite{Cuccaro2004, Gidney2018}, increasing resource requirements and potentially exceeding the capabilities of near-term devices~\cite{KanNam2021, Davoudi2025, Crane2024Hybrid}, although the overhead depends on the encoding and algorithm~\cite{Simon:2025pbo}. Preparing interacting initial states poses a further challenge~\cite{Gupta2025}.

Qubits are not the only building blocks for quantum computation. Some problems are more naturally addressed using $d$-level systems, or qudits~\cite{Bullock:2004oum,Rico:2018pas}. A trapped-ion qudit processor has been used to study a two-dimensional lattice gauge theory with tunable gauge-field truncations~\cite{Meth:2023wzd}, and large-spin lattice models have been proposed to regularise scalar field theories with controlled continuum extrapolations~\cite{Calliari:2024ppe}. In addition, physical oscillators offer a natural encoding for bosonic fields, with an unbounded Fock space. Continuous-variable quantum computers exploit this~\cite{Weedbrook2012,Marshall:2015mna,Abel:2024kuv,Abel:2025zxb}, and measurement-based photonic processors have been proposed for computing scattering amplitudes~\cite{Briceno:2023xcm}. However, these devices have no native representation for fermions. Fermions must either be bosonised, which requires additional gauge fields in local lattice constructions beyond one spatial dimension~\cite{Chen:2017fvr, Chen:2018nog, Chen:2019wlx}, or be mapped onto qubits with a Jordan--Wigner transformation, with each qubit encoded, for example, in a pair of modes sharing one photon (dual-rail)~\cite{Chuang:1995sm}. In linear-optical implementations, entangling gates between photonic qubits are probabilistic and require heralding and feed-forward~\cite{Knill_2001}.

In this study, we explore trapped-ion quantum devices as a platform for simulating resonant scattering and decay in theories where fermions interact with bosons. The internal states of a trapped ion form a qubit, while the collective motional modes of an ion chain form physical harmonic oscillators, or qumodes~\cite{Leibfried2003,Bruzewicz2019}. A single device can therefore hold fermions on qubits and bosons on oscillators. The two registers are coupled by spin-dependent forces, which displace a motional mode conditionally on the qubit state of the driven ion. In this way, any coupling linear in a bosonic field can be implemented using native operations~\cite{DavoudiLinkePagano2021,Liu2024ISA,Crane2024Hybrid}. Field theories with this structure can be simulated on trapped-ion quantum computers~\cite{Araz2024HybridTI}, and superconducting circuits, which couple qubits to microwave cavities, offer the same hybrid structure~\cite{Blais2021,Schmidt:2012za}.

We apply this idea to the $(1+1)$-dimensional Yukawa model, a Dirac fermion coupled to a real scalar, whose interaction is linear in the scalar and therefore native to a trapped-ion device. Trapped-ion experiments have already begun to explore Yukawa dynamics, demonstrating quench evolution in a Yukawa-type lattice model~\cite{Than2025} and the time evolution of the static Yukawa model~\cite{Wang:2026kcu}. We study two processes: the decay of an unstable scalar particle into a fermion--antifermion pair, and the collision of a fermion with an antifermion, in which the scalar particle forms briefly as a resonance. We design both for a trapped-ion device, emulate the device classically with tensor networks~\cite{Schollwock2011,Paeckel2019}, and compare what it would measure with perturbation theory. For the couplings we consider, the fitted scalar decay rate is $2$--$8\%$ below that of the Dyson-resummed one-loop evolution and is consistent with the next-to-leading-order result. At the larger coupling, the resonance peak also lies at a lower collision energy than predicted by the one-loop approximation, consistent with higher-order corrections. Resolving these differences requires control of the bosonic truncation and, in our emulations, six Fock levels per site oscillator give converged results.

The remainder of this paper is structured as follows. Section~\ref{sec:yuk} introduces the lattice Yukawa theory, its free particles and scalar normal modes, and the lattice perturbation theory against which we compare. Section~\ref{sec:device} maps the lattice model onto a trapped-ion register, converts its time evolution into native gates, and describes how the initial states are prepared and the observables measured. Section~\ref{sec:mps} describes the classical emulation of the device with MPS. Section~\ref{sec:results} presents results on the decay of a single scalar and on fermion--antifermion scattering. The conclusions are in Section~\ref{sec:conclusion}.
% ----------------------------

\section{Lattice Yukawa theory}
\label{sec:yuk}

%%%%%%%%%%%%%%%%%%%%%%%%%%%%%%%%
\subsection{Continuum and lattice Hamiltonian}
\label{sec:lattice}

The continuum Yukawa Hamiltonian is
\begin{align}
H=\int\!\mathrm{d}x\,\Big[&\bar\psi\big(-i\gamma^{1}\partial_{x}+m_{\psi}\big)\psi+g\,\varphi\,\bar\psi\psi\nonumber\\
&+\tfrac12\Pi^{2}+\tfrac12(\partial_{x}\varphi)^{2}+\tfrac12m_{\varphi}^{2}\varphi^{2}\Big],
\label{eq:Hcont}
\end{align}
where $\psi$ is a two-component Dirac field of mass $m_{\psi}$ and $\varphi$ is a real scalar of mass $m_{\varphi}$ with conjugate momentum $\Pi$. The Dirac matrices are $\gamma^{0}=\sigma^{z}$ and $\gamma^{1}=i\sigma^{y}$, and $\bar{\psi}=\psi^{\dagger}\gamma^{0}$. In $1+1$ dimensions, $\varphi$ is dimensionless, and $g$ has mass dimension. We discretise Eq.~\eqref{eq:Hcont} on an open chain of $N$ sites with spacing $b$. To avoid fermion doubling, we use Kogut--Susskind staggering, with one fermionic mode $\psi_{j}$ on each of the $N$ sites, satisfying $\{\psi_{i},\psi_{j}^{\dagger}\}=\delta_{ij}$. Grouping the sites into pairs gives $N/2$ physical sites, and the two modes in each physical site supply the two components of a Dirac spinor~\cite{Kogut:1974ag, Susskind:1976jm, Banks:1975gq}. Each site carries a scalar amplitude $\varphi_{j}$ and a momentum $\Pi_{j}$. The lattice Hamiltonian is $H=\Hf+\Hs+\Hint$, with
\begin{align}
\Hf&=\frac{i}{2b}\sum_{j=1}^{N-1}\big(\psi^{\dagger}_{j}\psi_{j+1}-\psi^{\dagger}_{j+1}\psi_{j}\big)
+m_{\psi}\sum_{j=1}^{N}(-1)^{j}\psi^{\dagger}_{j}\psi_{j},
\label{eq:Hf}\\
\Hs&= b\sum_{j=1}^{N}\left[\frac{\Pi_{j}^{2}}{2}+ \frac{m_{\varphi}^{2}}{2}\,\varphi_{j}^{2}\right]+\sum_{j=0}^{N}\frac{(\varphi_{j+1}-\varphi_{j})^{2}}{2b},\label{eq:Hs}\\
\Hint&=g\sum_{j=1}^{N}(-1)^{j}\,\varphi_{j}\,\psi^{\dagger}_{j}\psi_{j},
\label{eq:Hint}
\end{align}
where $\varphi_{0}=\varphi_{N+1}=0$ and the conjugate field obeys $[\varphi_{j},\varphi_{j'}] = [\Pi_{j},\Pi_{j'}] = 0$ and $[\varphi_{j},\Pi_{j'}]=(i/b)\,\delta_{jj'}$. 
%The staggered sign makes $(-1)^{j}\psi^{\dagger}_{j}\psi_{j}$ the lattice $\bar\psi\psi$, so 
Eq.~\eqref{eq:Hint} is the Yukawa vertex.
%which conserves the fermion number and creates and annihilates fermion--antifermion pairs. 
Since the masses and coupling enter the dimensionless lattice Hamiltonian only through the combinations $m_{\psi}b$, $m_{\varphi}b$ and $gb$, we express energies in units of $m_{\psi}$ and times in units of $1/m_{\psi}$, and choose the lattice spacing such that $m_{\psi}b=1$.

%%%%%%%%%%%%%%%%%%%%%%%%%%%%%%%%
\subsection{Free particles and scalar normal modes}
\label{sec:free}

The free part of the lattice Hamiltonian determines the particles we prepare and detect. At $g=0$, the single-particle spectrum consists of a positive-energy band and a negative-energy band, $\pm\omega(k)$, with
\begin{equation}
\omega(k)=\sqrt{\frac{\sin^{2}(kb)}{b^{2}}+m_{\psi}^{2}}\,.
\label{eq:disp-f}
\end{equation}
In the free vacuum, the negative-energy band is filled. A particle is an occupied state of the positive-energy band, and an antiparticle is a hole in the negative-energy band.

For the scalar, we introduce local oscillator operators $a_{j}$ at each site and write
\begin{equation}
\varphi_{j}=\frac{a_{j}+a_{j}^{\dagger}}{\sqrt{2m_{\varphi}b}},\qquad \Pi_{j}=i\sqrt{\frac{m_{\varphi}}{2b}}\,\bigl(a_{j}^{\dagger}-a_{j}\bigr),
\label{eq:siteosc}
\end{equation}
where $[a_{j},a_{j'}^{\dagger}]=\delta_{jj'}$, so that $a_{j}^{\dagger}$ creates a scalar quantum localised on site $j$. The site oscillators are not the normal modes of the free scalar, because the gradient term in Eq.~\eqref{eq:Hs} couples neighbouring sites. On the open chain, the normal modes are the standing waves $U_{jn}=\sqrt{2/(N+1)}\sin(p_{n}jb)$, with momenta $p_{n}=\pi n/[(N+1)b]$ and frequencies $\varepsilon_{n}\equiv\varepsilon(p_{n})$, where
\begin{equation}
\varepsilon(p)=\sqrt{m_{\varphi}^{2}+\frac{4}{b^{2}}\sin^{2}\frac{pb}{2}}\,.
\label{eq:disp}
\end{equation}
Both bands, Eqs.~\eqref{eq:disp-f} and~\eqref{eq:disp}, reduce to the continuum dispersions as \mbox{$b\to0$}. With the ladder operators $d_{n}$ of the normal modes, $[d_{n},d_{n'}^{\dagger}]=\delta_{nn'}$, the free scalar is diagonal,
\begin{equation}
\Hs=\sum_{n=1}^{N}\varepsilon_{n}\bigl(d_{n}^{\dagger}d_{n}+\tfrac12\bigr),
\label{eq:Hsmodes}
\end{equation}
and for open boundaries the two sets of operators are related by a sine transform and a single-mode squeeze,
\begin{eqnarray}
a_{j}&=&\sum_{n=1}^{N}U_{jn}\bigl(\cosh r_{n}\,d_{n}-\sinh r_{n}\,d_{n}^{\dagger}\bigr),\nonumber\\ r_{n}&=&\tfrac12\ln\frac{\varepsilon_{n}}{m_{\varphi}}\ .
\label{eq:aj-dn}
\end{eqnarray}
The field at site $j$ therefore involves every mode,
\begin{equation}
\varphi_{j}=\sum_{n=1}^{N}\frac{U_{jn}}{\sqrt{2b\varepsilon_{n}}}\bigl(d_{n}+d_{n}^{\dagger}\bigr)\ .
\label{eq:phi-dn}
\end{equation}
Both $a_j^\dagger$ ($a_j$) and $d_n^\dagger$ ($d_n$) create (annihilate) scalar quanta, but in different bases. The operator $a_j^\dagger$ adds a quantum to the oscillator at site $j$, while $d_n^\dagger$ adds one to the $n$th normal mode; a standing wave extending across the chain. Because $r_n\neq 0$, the two bases have different vacua. In particular, the ground state of $\Hs$, annihilated by every $d_n$, is a squeezed state in the site-oscillator basis. %This distinction also reflects their roles in our study. We use site oscillators for the tensor-network emulation in Sec.~\ref{sec:mps} and normal modes for the trapped-ion device in Sec.~\ref{sec:device}.

\subsection{Lattice perturbation theory}
\label{sec:pt}

We compare the emulation with perturbation theory of the lattice model at the same lattice spacing. At leading order, a scalar at rest decays into a fermion--antifermion pair with momenta $\pm k^{*}$, where $2\omega(k^{*})=m_{\varphi}$, with the width \mbox{$\Gamma_{\rm LO}=g^{2}\beta/[2m_{\varphi}\cos(k^{*}b)]$}, where \mbox{$\beta=\sqrt{1-4m_{\psi}^{2}/m_{\varphi}^{2}}$}. Our main benchmark includes the Dyson summation of the one-loop scalar self-energy, the fermion--antifermion bubble. This provides the leading approximation across the entire kinematic range we explore. We denote by $M$ and $\Gamma_{\rm D}$ the mass and width of the resulting complex pole. The next-to-leading-order (NLO) pole width, $\Gamma_{\rm NLO}$, is obtained by computing the absorptive part of the two-loop scalar self-energy.

When considering the time evolution, the Dyson sum is implemented faithfully by approximating the Hamiltonian in the random phase approximation (RPA)~\cite{BohmPines1953, FetterWalecka1971}. This approximation simplifies the evolution sufficiently that it can be performed using a classical computer, for the same initial state and on the same open chain as the emulation. When calculating the reflection probability, we use the Dyson-resummed $s$-channel amplitude and also include the tree-level $t$-channel exchange amplitude, and we average the result over the momentum spread of the incoming packets.

%%%%%%%%%%%%%%%%%%%%%%%%%%%%%%%%%%%%%%%
\section{Simulation on a trapped-ion device}
\label{sec:device}

%%%%%%%%%%%%%%%%%%%%%%%%%%%%%%%%
\subsection{Hybrid register and Hamiltonian}
\label{sec:register}
\begin{figure}[t]
    \centering
    \begin{tikzpicture}[
      scale=1.35, line width=0.7pt, font=\small,
      ion/.style={draw, circle, minimum size=6.5mm,inner sep=0pt},
      even/.style={ion,fill=cgreen!25},
      odd/.style={ion,fill=cblue!25},
      lab/.style={font=\scriptsize}]
    
        \foreach \i/\x/\st/\nm in {1/0/odd/j, 2/1.35/even/j+1, 3/2.7/odd/j+2, 4/4.05/even/j+3}{
      \node[\st, minimum size=8.5mm] (q\i) at (\x,0) {};
      \node[lab, font=\small] at (\x,0) {$q_{\nm}$};
      \draw[-{Latex[length=1.3mm]}] (\x,-0.47)--(\x,-0.95);
      \node[lab,anchor=north, font=\small] at (\x,-1.0) {$\varphi_{\nm}$};
    }
    \foreach \a/\b in {1/2,2/3,3/4}{\draw[thick] (q\a)--(q\b);}
    \draw[thick,dashed] (q4) -- ++(0.75,0);
    \draw[thick,dashed] (q1) -- ++(-0.75,0);
    
    \draw[decorate,decoration={brace,amplitude=4pt}] (-0.45,0.55) -- (1.80,0.55)
      node[midway,above=4pt,lab]{physical site $\ell$};
    \draw[decorate,decoration={brace,amplitude=4pt}] (2.25,0.55) -- (4.50,0.55)
      node[midway,above=4pt,lab]{physical site $\ell+1$};
    
    \node[lab,anchor=east, font=\large] at (-0.85,0){$\psi$};
    \begin{scope}[yshift=-0.7mm]
      \foreach \x in {0,1.35,2.7,4.05}{
        \draw[domain=-0.321:0.321,smooth,variable=\u] plot ({\x+\u},{-1.84+5.24*\u*\u});
        \foreach \yl in {-1.72,-1.57,-1.39}{
          \draw ({\x-sqrt((\yl+1.84)/5.24)},\yl) -- ({\x+sqrt((\yl+1.84)/5.24)},\yl);}
      }
      \node[lab,anchor=east, font=\large] at (-0.85,-1.5){$\varphi$};
      \draw[|-|] (0,-2.15)--(1.35,-2.15);
      \node[lab,anchor=north, font=\large] at (0.675,-2.2){$b$};
    \end{scope}
    \end{tikzpicture}
    \caption{Site-basis representation of four consecutive staggered lattice sites, separated by spacing $b$. Each staggered site $j$ contains one fermionic mode, encoded by a Jordan--Wigner qubit $q_j$, and one site oscillator $a_j$ representing the scalar field $\varphi_j$. Colours distinguish the odd and even sublattices, and each bracketed pair forms a physical site $\ell$, with $j=2\ell-1$, whose two staggered sites hold the two components of a Dirac spinor. On a trapped-ion device, $q_j$ is encoded in the internal states of ion $j$, while the scalar is stored in collective motional modes $d_n$, related to the site oscillators by Eq.~\eqref{eq:aj-dn}.}
    \label{fig:lattice}
\end{figure}
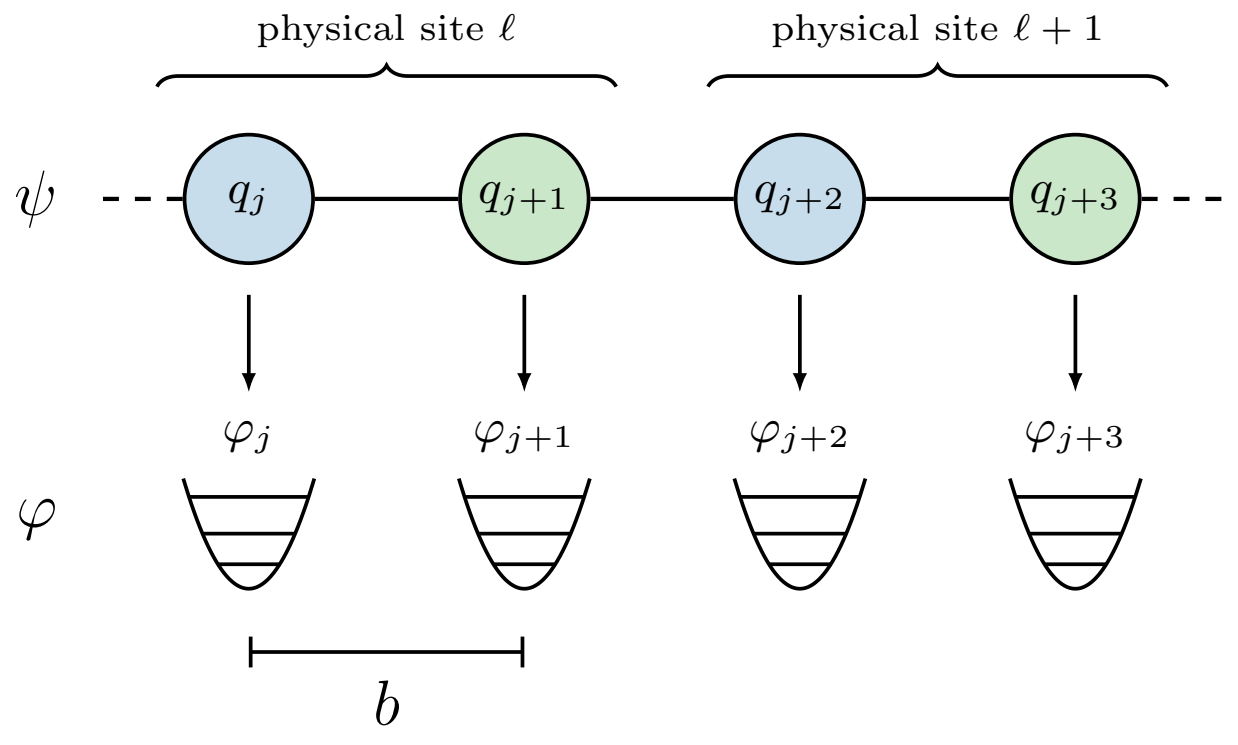

On a trapped-ion device, each staggered site is one ion, whose qubit holds the fermion through the Jordan--Wigner transformation~\cite{JordanWigner1928}. An ion chain also has collective motional modes, as many in each transverse direction as there are ions, and we hold the scalar in one mode per lattice site~\cite{DavoudiLinkePagano2021, Than2025}, i.e. motional mode $n$ holds the normal mode $d_{n}$ of Sec.~\ref{sec:free}. The quantum register is sketched in Fig.~\ref{fig:lattice}. The qubits $q_{j}$ of neighbouring ions are coupled by the hopping term of Eq.~\eqref{eq:Hf}, and each qubit couples through the Yukawa term of Eq.~\eqref{eq:Hint} to the scalar at its own site, $\varphi_{j}$, drawn as a harmonic oscillator with equally spaced energy levels. The sketch shows the site oscillators of Eq.~\eqref{eq:siteosc}, in which every coupling is local.

To represent the Yukawa Hamiltonian on the quantum device, one first needs to convert each term into device-native operators. After a Jordan--Wigner transformation, Eq.~\eqref{eq:Hf} can be written as
\begin{align}
\Hf^\prime&=\frac{1}{4b}\sum_{j=1}^{N-1}\bigl(\sx_{j}\sx_{j+1}+\sy_{j}\sy_{j+1}\bigr)\nonumber\\
&\quad+\frac{m_{\psi}}{2}\sum_{j=1}^{N}(-1)^{j}\sz_{j}+\text{const.}
\label{eq:Hfspin}
\end{align}
The free scalar Hamiltonian needs no rewriting, since the motional modes are its normal modes and Eq.~\eqref{eq:Hsmodes} is already in device-native form. Inserting Eq.~\eqref{eq:phi-dn} into Eq.~\eqref{eq:Hint} gives the interaction term
\begin{equation}
\Hint^\prime=\frac{g}{2\sqrt{2b}}\sum_{j,n=1}^{N}\frac{(-1)^{j}U_{jn}}{\sqrt{\varepsilon_{n}}}\bigl(\mathbb{I}_{j}+\sz_{j}\bigr)\bigl(d_{n}+d_{n}^{\dagger}\bigr).
\label{eq:Hintspin}
\end{equation}

%%%%%%%%%%%%%%%%%%%%%%%%%%%%%%%%
\subsection{Native gates and Trotter evolution}
\label{sec:gates}

The Trotter expansion of a Hamiltonian for a time step  $\delta t$ is
\begin{align}
e^{-iH\delta t}&\simeq e^{-i\Hf^\prime\delta t/2}\,e^{-i\Hs\delta t/2}\,e^{-i\Hint^\prime\delta t}\nonumber\\
&\quad\times e^{-i\Hs\delta t/2}\,e^{-i\Hf^\prime\delta t/2},
\label{eq:strang}
\end{align}
with an error of order $\delta t^{3}$ per step. Each term in our Hamiltonian can be implemented using native trapped-ion operations~\cite{Araz2024HybridTI}.
%In Eq.~\eqref{eq:Hfspin}, the hopping is a brickwork of nearest-neighbour $XX{+}YY$ rotations on alternating bonds.
In Eq.~\eqref{eq:Hfspin}, the hopping terms on even and odd bonds do not commute, so each factor $e^{-i\Hf^\prime\delta t/2}$ is itself split symmetrically into a brickwork of nearest-neighbour $XX{+}YY$ rotations on alternating bonds, which keeps the error of order $\delta t^{3}$ per step. Since $\sx_{j}\sx_{j+1}$ and $\sy_{j}\sy_{j+1}$ commute, each rotation is two M\o{}lmer--S\o{}rensen gates~\cite{Molmer1999,Solano1999}, the second conjugated by single-qubit rotations. The staggered mass is a layer of single-qubit $z$ rotations with alternating sign, applied virtually as phase shifts of the subsequent drives. Over a Trotter step of duration $\delta t$, Eq.~\eqref{eq:Hsmodes} generates a phase rotation $e^{-i\varepsilon_{n}\delta t\,d_{n}^{\dagger}d_{n}}$ of each motional mode. In Eq.~\eqref{eq:Hintspin}, the $\sz_{j}$ term exponentiates to a displacement of mode $n$ conditioned on qubit $j$, $\mathrm{CD}_{jn}(z)=\exp[\sz_{j}(zd_{n}^{\dagger}-z^{*}d_{n})]$, which is the spin-dependent force of a trapped-ion device, driven by simultaneous red- and blue-sideband tones~\cite{Leibfried2003, Araz2024HybridTI}. The $\mathbb{I}_{j}$ term exponentiates to an unconditional displacement $D_{n}(z)=\exp(zd_{n}^{\dagger}-z^{*}d_{n})$, which acts only on the even modes, because $\sum_{j}(-1)^{j}U_{jn}$ vanishes for odd $n$ on a chain with even $N$. The staggered sign and the standing wave enter only through the amplitudes, $z_{jn}\propto(-1)^{j}U_{jn}/\sqrt{\varepsilon_{n}}$. All terms of Eq.~\eqref{eq:Hintspin} are products of $\sz_{j}$ or $\mathbb{I}_{j}$ with the quadratures $d_{n}+d_{n}^{\dagger}$, so they commute, and one Trotter step of the interaction is $N^{2}$ conditional and $N/2$ unconditional displacements in any order. The M\o{}lmer--S\o{}rensen gates of the hopping layer can be driven through transverse or axial modes other than those that hold the scalar, so they leave it undisturbed~\cite{Araz2024HybridTI}. No gate depends on a Fock cutoff, whereas encoding the scalar in qubits in the binary harmonic-oscillator basis costs $\mathcal{O}(\log^{2}\Lambda)$ entangling gates per interaction term~\cite{DavoudiLinkePagano2021}. 
We refer the reader to Refs.~\cite{Liu2024ISA, Crane2024Hybrid} for the details of decomposition of each layer into native gates.

%%%%%%%%%%%%%%%%%%%%%%%%%%%%%%%%
\subsection{State preparation}
\label{sec:prep}

Each simulation begins by preparing the vacuum $\ket{\Omega}$ of the interacting Hamiltonian. On the device, it would be prepared adiabatically~\cite{JordanLeePreskill2014} or variationally~\cite{Peruzzo2014, Cerezo2021, Araz2024HybridTI, Zemlevskiy:2024vxt}. In the emulation, however, the vacuum is prepared via imaginary-time evolution, see Sec.~\ref{sec:mps}.

We construct fermion wave packets with Gaussian momentum envelopes using the bands of the free fermion Hamiltonian. The particle packet is created by adding a fermion to the positive-energy band, and the antiparticle packet by removing one from the negative-energy band. These packet operators act on the interacting vacuum $\ket{\Omega}$, with the associated rotations implemented through a Givens network~\cite{Chai:2023qpq}, and create fermionic excitations without their full scalar dressing. This dressing develops during the evolution and can generate transient scalar radiation. The incoherent observables defined below remove the coherent contribution from the dressing and radiation, while retaining their fluctuations.

To study the decay, we prepare a single scalar quantum at rest at site $x_{0}$. We build it as a wave packet of the motional modes, $A_{f}^{\dagger}=\sum_{n}\tilde f_{n}d_{n}^{\dagger}$, whose amplitudes $\tilde f_{n}\propto e^{-p_{n}^{2}/4\sigma_{p}^{2}}U_{x_{0}n}$ combine the standing waves into a lump centred at $x_{0}$, with momenta spread by $\sigma_{p}$ around zero. However, acting with $A_{f}^{\dagger}$ on $\ket{\Omega}$ alone is not sufficient. The fermion vacuum carries a nonzero $\bar\psi\psi$, which the Yukawa term turns into a static source for the scalar, so that $\langle d_{n}\rangle_{\Omega}\neq0$ and $A_{f}^{\dagger}\ket{\Omega}$ still overlaps with the vacuum. We remove this overlap,
\begin{equation}
\ket{\varphi_{f}}\propto\bigl(A_{f}^{\dagger}-\bar\alpha^{*}\bigr)\ket{\Omega},\qquad \bar\alpha=\bra{\Omega}A_{f}\ket{\Omega},
\label{eq:fock}
\end{equation}
so that the state holds one quantum on top of the background field and has no vacuum component, \mbox{$\braket{\Omega|\varphi_{f}}=0$}.\footnote{Similar approaches have been taken in the literature; see Refs.~\cite{Turco:2023rmx, Davoudi:2024wyv}.} On the device, this takes three steps, because \mbox{$A_{f}^{\dagger}-\bar\alpha^{*}=D_{f}(\bar\alpha)A_{f}^{\dagger}D_{f}(\bar\alpha)^{\dagger}$}~\cite{de_Oliveira_1990}: a displacement of the packet mode by $-\bar\alpha$ removes the background field, a blue-sideband transition on an ancilla ion adds one phonon~\cite{Leibfried2003}, and the opposite displacement restores the background. 

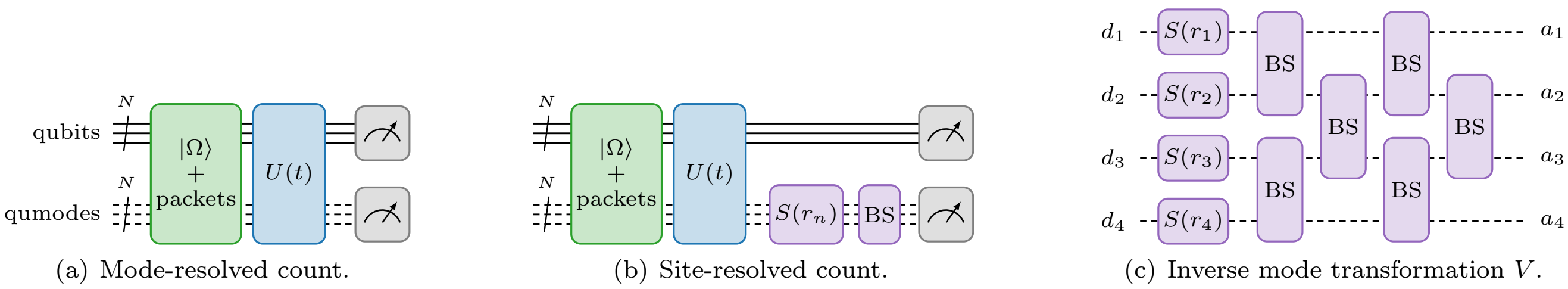
\begin{figure*}[!t]
\centering
\tikzset{
  qw/.style={line width=0.55pt},
  mw/.style={line width=0.55pt,dash pattern=on 2pt off 1.5pt},
  blk/.style={rounded corners=3pt,inner sep=1.5pt,font=\scriptsize,align=center,line width=0.6pt},
  prep/.style={blk,draw=cgreen,fill=cgreen!25},
  evo/.style={blk,draw=cblue,fill=cblue!25},
  gauss/.style={blk,draw=cpurple,fill=cpurple!25},
  rot/.style={blk,draw=cred!80!black,fill=cred!35},
  wl/.style={font=\scriptsize,anchor=east},
  pics/meter/.style={code={
    \node[blk,draw=cgrey,fill=cgrey!25,minimum width=6mm,minimum height=6mm] (-box) {};
    \draw[line width=0.5pt] ([xshift=-2.0mm,yshift=-0.9mm]-box.center) arc[start angle=160,end angle=20,radius=2.1mm];
    \draw[line width=0.5pt,-{Latex[length=1mm]}] ([yshift=-0.9mm]-box.center)--++(55:2.9mm);
  }},
}
% a register = three parallel lines at y +- 0.12 around its centre; qubits at y = 1, qumodes at y = 0
\newcommand{\bundle}[3]{\foreach \dy in {-0.12,0,0.12}{\draw[#1] (-9,#2+\dy)--(#3,#2+\dy);}}
\newcommand{\regs}[2]{%
  \bundle{qw}{1}{#1} \bundle{mw}{0}{#1}
  \ifnum#2=1 \node[wl] at (-9.5,1) {qubits}; \node[wl] at (-9.5,0) {qumodes};\fi
  \draw[line width=0.45pt] (-5.7,0.78)--(-4.3,1.22); \node[font=\tiny,above] at (-5,1.2) {$N$};
  \draw[line width=0.45pt] (-5.7,-0.22)--(-4.3,0.22); \node[font=\tiny,above] at (-5,0.2) {$N$};
  \node[prep,minimum height=15.5mm,minimum width=9mm] at (16,0.5) {$\ket{\Omega}$\\$+$\\packets};
  \node[evo,minimum height=15.5mm,minimum width=8mm] at (44,0.5) {$U(t)$};
  \pic at (#1,1) {meter}; \pic at (#1,0) {meter};
}
\subfigure[Mode-resolved count.\label{fig:circ-mode}]{%
\begin{tikzpicture}[x=0.37mm,y=9mm]
  \regs{72}{1}
\end{tikzpicture}}\qquad\qquad
\subfigure[Site-resolved count.\label{fig:circ-gauss}]{%
\begin{tikzpicture}[x=0.37mm,y=9mm]
  \regs{115}{0}
  \node[gauss,minimum height=6.5mm] at (73,0) {$S(r_{n})$};
  \node[gauss,minimum height=6.5mm] at (95,0) {BS};
\end{tikzpicture}}\qquad\qquad
\subfigure[Inverse mode transformation $V$.\label{fig:circ-V}]{%
\begin{tikzpicture}[x=0.37mm,y=7mm]
  \foreach \k in {1,...,4}{\draw[mw] (-4,-\k)--(112,-\k); \node[wl] at (-5,-\k) {$d_{\k}$}; \node[font=\scriptsize,anchor=west] at (113,-\k) {$a_{\k}$};}
  \foreach \k in {1,...,4}{\node[gauss,minimum height=5mm] at (12,-\k) {$S(r_{\k})$};}
  \foreach \x/\a in {38/1,38/3,57/2,76/1,76/3,95/2}{%
    \node[gauss,minimum height=11.5mm,minimum width=5mm] at (\x,-\a-0.5) {BS};}
\end{tikzpicture}}
\caption{Circuits for the two kinds of scalar measurement. State and wave packet preparation is shown in green, time evolution in blue and measurements in grey; solid lines represent qubits and dashed lines the motional modes (qumodes). (a)~Mode-resolved count: each motional mode is measured directly, together with the qubits. (b)~Site-resolved count: the modes are first converted into the site oscillators (purple) and then measured. The purple set of gates is expanded for four motional modes in panel~(c) with a single-mode squeezer gate, $S(r_{n})$, on each mode followed by a mesh of beam splitters (BS).}
\label{fig:readout-circuits}
\end{figure*}

%%%%%%%%%%%%%%%%%%%%%%%%%%%%%%%%
\subsection{Observables and readout}
\label{sec:obs}

We track both processes through the fermion and scalar sectors. For the fermions, we examine the charge flow and reflection probability, while the scalar observables reveal the mediator and its decay. We measure all observables relative to the vacuum, and the charge and scalar observables are constructed from the native measurements shown in Fig.~\ref{fig:readout-circuits}. 

Projective $\sz_j$ measurements determine the site occupations. Subtracting their expectation values in the interacting vacuum gives the excess charge density,
\begin{equation}
\delta\langle n_{j}\rangle=\langle\psi^{\dagger}_{j}\psi_{j}\rangle-\langle\psi^{\dagger}_{j}\psi_{j}\rangle_{\Omega}\, .
\label{eq:dn}
\end{equation}
However, this observable is not the natural quantity to follow in the decay of a stationary scalar. The fermion--antifermion pair is produced in a superposition of opposite-charge configurations, so each track carries no net charge and the daughters appear in $\delta\langle n_{j}\rangle$ only as an alternation between even and odd sites. Instead, we consider the excess staggered density
\begin{equation}
\delta\rho_{j}=(-1)^{j}\delta\langle n_{j}\rangle,
\label{eq:drho}
\end{equation}
which removes this alternation and corresponds to the density operator coupled to $\varphi_{j}$.
The qubit number operator is native to the trapped-ion device and can be measured by a simple projective measurement, as shown in Fig.~\ref{fig:circ-mode}.

The scattering probability can be inferred from the reflection of localised charge. For a particle entering from the left and an antiparticle from the right, the charge to the right of the collision centre $x_c$, $Q_R(t)=\sum_{j>x_c}\delta\langle n_j\rangle$, evolves from $-1$ before the collision to $1-2R$ once the outgoing packets have separated. The reflection probability therefore follows as
\begin{equation}
R(t)=\tfrac12\bigl[1-Q_R(t)\bigr].
\label{eq:Rcharge}
\end{equation}
In one spatial dimension, scattering occurs only forwards or backwards, so  $R(t\to\infty)$ gives the backward-scattering probability. Its time evolution reveals the collision dynamics, while its dependence on $\sqrt{s}$ maps the resonance lineshape.

% The charge observables do not reveal whether the collision proceeds through an on-resonance scalar mediator. 
The charge observables alone do not directly resolve the scalar excitation associated with the resonance. To probe its production and decay, we measure the incoherent occupation of each scalar motional mode. 
% The total occupation $\langle d_n^\dagger d_n\rangle$ includes both quantum excitations and the contribution from the mode's coherent amplitude, which can arise from the vacuum condensate or the scalar clouds accompanying the fermions and the radiation emitted as they form. 
The total occupation $\langle d_n^\dagger d_n\rangle$ is the sum of the coherent contribution $|\langle d_n\rangle|^2$ and the occupation associated with fluctuations about the mean. The coherent amplitude can receive contributions from the vacuum condensate, the scalar clouds accompanying the fermions, and radiation emitted as the clouds form.
Subtracting the squared coherent amplitude isolates the incoherent occupation. Subtracting its vacuum value then removes the background present before the collision,
\begin{equation}
n_{\rm inc}(p_{n},t)=\langle d_{n}^{\dagger}d_{n}\rangle-|\langle d_{n}\rangle|^{2}-\Bigl(\langle d_{n}^{\dagger}d_{n}\rangle_{\Omega}-|\langle d_{n}\rangle_{\Omega}|^{2}\Bigr).
\label{eq:ninc-pn}
\end{equation}
Each mode's incoherent occupation requires three native measurements: the average phonon number ($\langle d_{n}^{\dagger}d_{n}\rangle=\langle\hat N_{n}\rangle$) and the two homodyne quadratures ($\langle d_{n}\rangle=(\langle\hat X_{n}\rangle+i\langle\hat P_{n}\rangle)/\sqrt2$), with the circuit shown in Fig.~\ref{fig:circ-mode}~\cite{Araz2024HybridTI}. 
% Because a free motional mode retains its occupation under free evolution, this observable tracks interaction-induced changes without confusing them with the subsequent spatial spreading of the scalar field. 
Free scalar evolution conserves the occupation of each normal mode, so changes in this observable reflect interactions rather than the spatial spreading of a wave packet.
A scalar mediator produced nearly at rest preferentially populates the lowest-momentum modes, so the rise and fall of their incoherent occupation tracks its production and decay, allowing comparison with perturbation theory.

The mode-resolved occupation reveals which motional modes are populated, but not where the scalar excitations are located. To resolve their spatial dynamics, we measure the incoherent occupation of each site oscillator,
\begin{equation}
n_{\rm inc}(j)=\langle a_{j}^{\dagger}a_{j}\rangle-|\langle a_{j}\rangle|^{2}-\Bigl(\langle a_{j}^{\dagger}a_{j}\rangle_{\Omega}-|\langle a_{j}\rangle_{\Omega}|^{2}\Bigr),
\label{eq:ninc-j}
\end{equation}
which shows the mediator forming where the incoming packets overlap and, during decay, the parent excitation disappearing as its daughters propagate away. Unlike the motional modes $d_n$, the site oscillators $a_j$ are not directly accessible to the native readout. We therefore apply a Gaussian unitary $V$ satisfying $V^{\dagger}d_{j}V=a_{j}$, which maps the site-oscillator observables onto motional-mode observables that can be measured using the same readout as before~\cite{Weedbrook2012} (Fig.~\ref{fig:circ-gauss}). As shown in Fig.~\ref{fig:circ-V}, the transformation $V$ follows from Eq.~\eqref{eq:aj-dn} in two steps. A single-mode squeezer $S(r_{n})$ on each motional mode first undoes the squeeze in Eq.~\eqref{eq:aj-dn}. A mesh of two-mode beam splitters then applies the sine transform $U_{jn}$, which mixes the modes without changing the number of quanta. Any such $N$-mode transformation decomposes into $N(N-1)/2$ beam splitters~\cite{Reck1994}, which can be arranged in $N$ layers~\cite{Clements2016}. Meshes of this kind have been programmed to implement arbitrary multimode transformations on photonic chips~\cite{Carolan:2015yod}, including the closely related discrete Fourier transform~\cite{Crespi:2015jzv}, and on up to four motional modes of trapped ions~\cite{Chen:2022auh}, where squeezing~\cite{Meekhof1996, Burd2019} and beam splitters between motional modes~\cite{Gan2020, Hou2024} are native operations. The transformation requires $\mathcal{O}(N^{2})$ two-mode gates, making $n_{\rm inc}(j)$, Eq.~\eqref{eq:ninc-j}, the most resource-intensive observable considered here. For quantitative comparison with perturbation theory, we therefore use the mode-resolved occupation, Eq.~\eqref{eq:ninc-pn}, which also gives the more reliable count of scalar quanta without an ad hoc choice of spatial window, and we use Eq.~\eqref{eq:ninc-j} to visualise the dynamics of the unstable scalar.

After the collision, the system is generally in a coherent superposition of transmitted and reflected components, generating entanglement between the regions on either side. We quantify this using the vacuum-subtracted bipartite entanglement entropy between sites $j$ and $j+1$, $\Delta S_j=S_j-S_j^\Omega$, where $S_{j}=-\mathrm{Tr}\,\rho_{\le j}\log_{2}\rho_{\le j}$, with $\rho_{\le j}$ the reduced density matrix of the sites up to~$j$. In the MPS emulation of Sec.~\ref{sec:mps}, we compute the von Neumann entropy from the Schmidt values of the bonds. On hardware, extracting this quantity would require MPS tomography or randomised measurements~\cite{Elben2023}, techniques that have been demonstrated on trapped-ion qubits~\cite{Lanyon2017, Brydges2019}. Extending these methods to the qumodes is beyond the scope of this study.

%%%%%%%%%%%%%%%%%%%%%%%%%%%%%%%%
\section{Classical emulation with tensor networks}
\label{sec:mps}

Since the system size is beyond the reach of exact state-vector simulation and exact diagonalisation, and scalable trapped-ion hardware is not yet available to us, we emulate the device using a matrix product state (MPS)~\cite{Schollwock2011, Orus2014, Paeckel2019}. Each tensor holds one staggered site, the qubit, and a site oscillator $a_{j}$ truncated to its lowest $\Lambda$ Fock states, so the physical index has dimension $d=2\Lambda$. Neighbouring tensors share a bond of dimension at most $\chi$. We work with the site oscillators $a_{j}$, and every term in the Yukawa Hamiltonian acts on at most two neighbouring sites. We therefore use Eq.~\eqref{eq:aj-dn} to express the collective-mode dynamics and observables in the local $a_j$ basis used by the MPS, imposing the Fock cutoff on the site oscillators. The cutoff enters through the Rayleigh--Ritz projection $H_{\Lambda}=P_{\Lambda}HP_{\Lambda}$ onto the retained Fock states~\cite{Ritz_1909, MacDonald_1933}, with $\varphi_{j}^{2}$ and $\Pi_{j}^{2}$ formed before the projection. %Truncating $\varphi_{j}$ and $\Pi_{j}$ first lowers the energy of the top retained level and, at the smallest cutoffs, closes the free scalar gap.
This keeps every retained Fock level of a site oscillator at its exact energy, $m_{\varphi}(n+\tfrac12)$ for $n=0,\dots,\Lambda-1$. Squaring the truncated $\varphi_{j}$ and $\Pi_{j}$ instead would lower the highest retained Fock level, $n=\Lambda-1$, to $m_{\varphi}(\Lambda-1)/2$, because its coupling to the discarded level $n=\Lambda$ is lost. At $\Lambda=2$, it would then be degenerate with the vacuum, and the scalar would have no gap.

Real-time evolution is achieved by second-order time-evolving block decimation (TEBD)~\cite{Vidal2003, Vidal2004}, %which splits $H_{\Lambda}$ into even and odd bonds instead of the field terms used on the device. 
which groups the terms in $H_{\Lambda}$ by even and odd lattice bonds, whereas the device circuit applies separate Trotter layers for fermion dynamics on the qubits, free scalar dynamics on the qumodes, and the Yukawa interaction coupling the two registers.
Both Trotter decompositions are second order in the time step. When applied to the same Hamiltonian, they converge to the same evolution as $\delta t\to0$, and agreement with the untruncated device dynamics additionally requires convergence in the bond dimension $\chi$ and the Fock cutoff $\Lambda$. We obtain the interacting vacuum $\ket{\Omega}$ by imaginary-time TEBD from the half-filled staggered product state, annealing the step until the energy is stationary. We assess the numerical accuracy by changing $\Lambda$, $\chi$ and the Trotter step; where uncertainty bands are shown, these changes are added in quadrature.

%The emulation follows the device protocol of Sec.~\ref{sec:device} step by step, with three differences of implementation. We apply the Givens network of the fermion packets gate by gate. For the scalar quantum, we apply the operator of Eq.~\eqref{eq:fock} directly to the state and renormalise it, which prepares the ideal state without the ancilla.
The emulation follows the device protocol in Sec.~\ref{sec:device}, with three implementation differences. The time evolution is grouped by lattice bonds rather than field terms, as described above; the vacuum is prepared by imaginary-time evolution; and for the scalar quantum, we apply the operator in Eq.~\eqref{eq:fock} directly to the state and renormalise it, which prepares the ideal state without the ancilla.

%%%%%%%%%%%%%%%%%%%%%%%%%%%%%%%%
\section{Results}
\label{sec:results}
%%%%%%%%%%%%%%%%%%%%%%%%%%%%%%%%

\begin{figure*}[!th]
    \centering
    \includegraphics[width=0.95\linewidth]{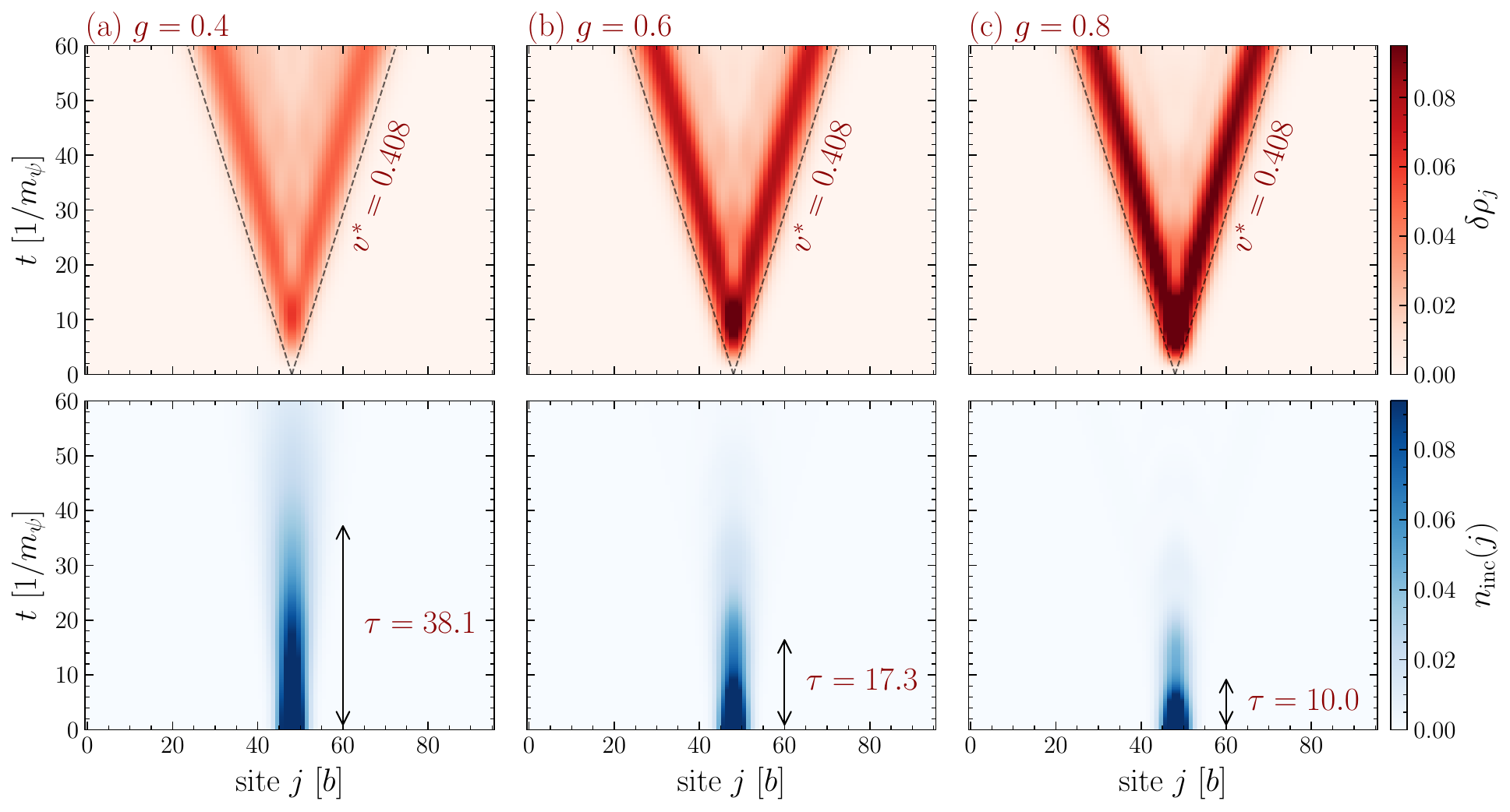}
    \caption{Decay of a single scalar quantum at rest at $x_{0}=48$ on a chain of $N=96$ sites ($\sigma_{p}=0.25$, $\Lambda=8$, $\chi=64$): (a)~$g=0.4$, (b)~$g=0.6$ and (c)~$g=0.8$. Upper row: the staggered density $\delta\rho_{j}$, which shows the fermion--antifermion pair produced by the decay. The dashed lines $x_{0}\pm v^{*}t$ mark daughters emitted at $t=0$ with the group velocity $v^{*}=0.408$. Lower row: the site-resolved scalar count $n_{\rm inc}(j)$, which shows the parent fading away; the double arrows mark its one-loop lifetime $\tau=1/\Gamma_D$. Adjacent sites are averaged.}
    \label{fig:decay-spacetime}
\end{figure*}

In this section, we present the emulated decay of a single scalar and fermion--antifermion scattering. For comparison with perturbation theory, we focus on the decay width $\Gamma$ of the scalar, the time dependence of the incoherent occupation of the slow motional modes, Eq.~\eqref{eq:ninc-pn}, and the reflection probability $R$.

%%%%%%%%%%%%%%%%%%%%%%%%%%%%%%%%
\subsection{Scalar decay} 

We begin with the decay of one scalar quantum at rest, $\varphi\to X$. We prepare it using Eq.~\eqref{eq:fock} and place it at the centre, $x_{0}=48$, of a chain of $N=96$ sites, with $\sigma_{p}=0.25$ and the bare mass $m_{\varphi}=2\sqrt{1.5}$. In the case of the dominant decay, $X=\psi \bar{\psi}$ and the daughters each leave with momentum $k^{*}=\pi/4$ and group velocity $v^{*}=0.408$. The MPS evolution is performed with $\chi = 64$, and we vary $\Lambda$. We evolve to $t=100$, which is before the daughters reach the ends of the chain.  

Figure~\ref{fig:decay-spacetime} shows the decay in space and time, up to \mbox{$t=60$}. The parent fades at $x_{0}$ while the staggered density of the daughters spreads along $x_{0}\pm v^{*}t$. At $g=0.4$, the parent lives longer, and the daughters are emitted over a longer time. This is why their density along the tracks is lower than at $g=0.8$.

\begin{table}[t]
\caption{Decay width of the scalar for different Fock cutoffs compared to perturbation theory. $\Gamma_{\rm D}$: one-loop Dyson pole width. $\Gamma_{\rm NLO}$: next-to-leading order pole width. Last five columns: the tensor network emulated width at Fock cutoff $\Lambda$, divided by the RPA width (both extracted by fitting to the decay, see text).}
\label{tab:decay-width}
\footnotesize
\begin{ruledtabular}
\begin{tabular}{ccc|ccccc}
 & & & \multicolumn{5}{c}{$\Gamma_{\rm TN}^{\rm fit}/\Gamma_{\rm RPA}^{\rm fit}$}\\
$g$ & $\Gamma_{\rm D}$ & $\Gamma_{\rm NLO}/\Gamma_{\rm D}$ & $\Lambda=2$ & $3$ & $4$ & $6$ & $8$\\
\colrule
$0.4$ & $0.0263$ & $0.986$ & $1.177$ & $1.073$ & $0.983$ & $0.979$ & $0.979$\\
$0.6$ & $0.0579$ & $0.967$ & $1.210$ & $1.050$ & $0.970$ & $0.966$ & $0.966$\\
$0.8$ & $0.0996$ & $0.940$ & $1.300$ & $1.027$ & $0.945$ & $0.939$ & $0.939$\\
\end{tabular}
\end{ruledtabular}
\end{table}

We extract the lifetime, $\tau$, by tracking the incoherent occupation of the motional modes containing the parent, $N_{\rm inc}(t)=\sum_{p_{n}\le3\sigma_{p}}n_{\rm inc}(p_{n},t)$. These modes contain $99.8\%$ of the initially prepared quantum, allowing us to follow its decay without imposing a spatial window. The decay is not purely exponential because the parent is not prepared in an exact eigenstate of the full Hamiltonian. We account for this by fitting $A\,e^{-\Gamma t}+C$~\cite{Jha2025} over $5\le t\le100$. The width obtained from the RPA fit overestimates the one-loop Dyson pole width by $3.7\%$, $3.1\%$ and $1.8\%$ at the three couplings. 

We summarise the results in Table~\ref{tab:decay-width}. For context, the leading-order width of Sec.~\ref{sec:pt} is $\Gamma_{\rm LO}=g^{2}/6$ for these parameters. The third column gives $\Gamma_{\rm NLO}/\Gamma_{\rm D}$. The remaining columns are the MPS emulations at different cutoffs, $\Lambda\in\{2,3,4,6,8\}$. For large enough $\Lambda$, we observe that the emulation is sensitive to physics beyond the RPA and consistent with the NLO correction.

%%%%%%%%%%%%%%%%%%%%%%%%%%%%%%%%
\subsection{Fermion--antifermion scattering} 

\begin{figure*}[thb]
    \centering
    \includegraphics[width=0.95\linewidth]{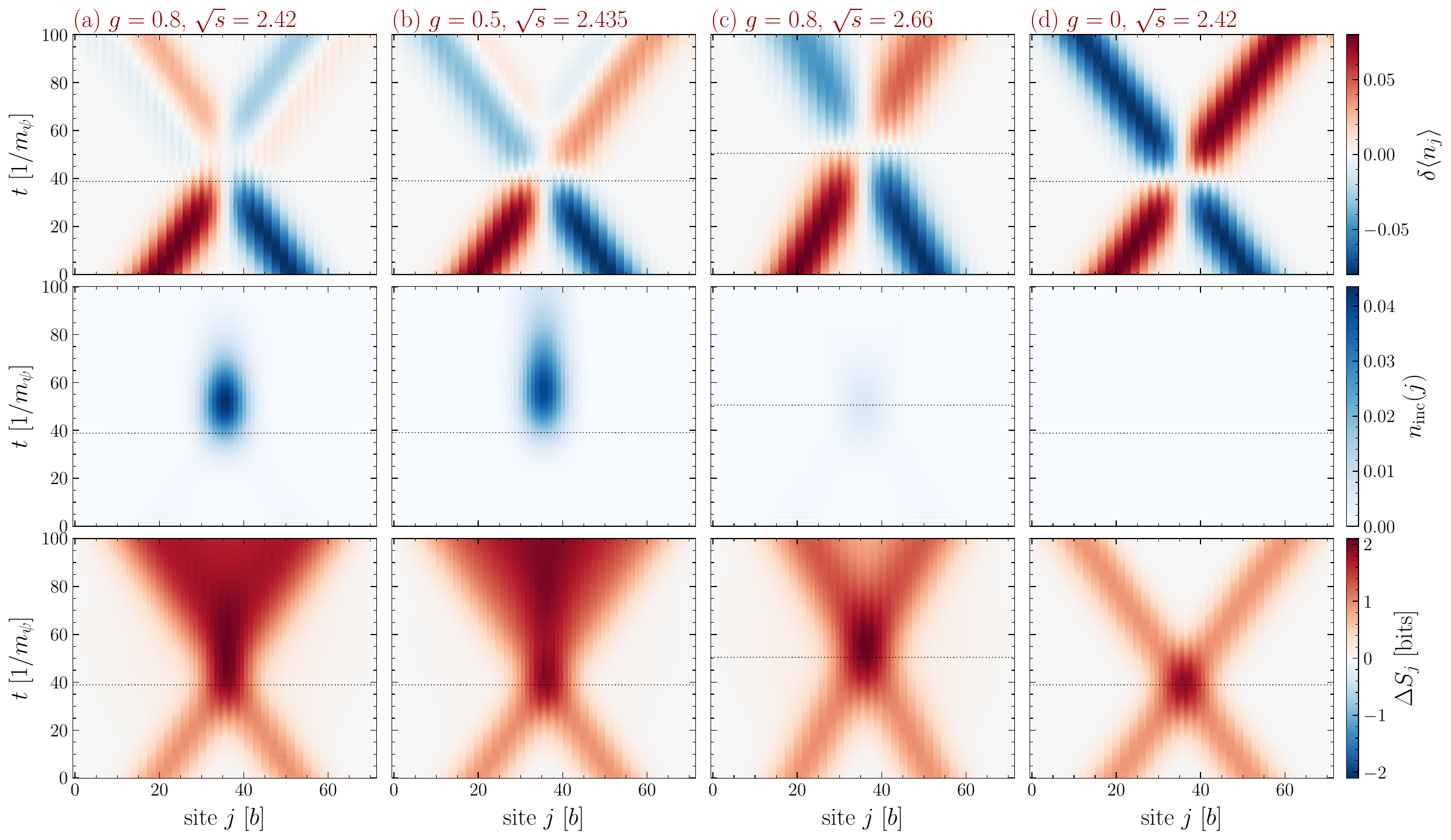}
    \caption{Space-time picture of fermion--antifermion collisions at $\Lambda=6$ ($N=72$, $\sigma_{k}=0.1$, $\chi=64$). Columns: (a)~on resonance, $g=0.8$, $\sqrt{s}=2.42$; (b)~a longer-lived mediator, $g=0.5$, $\sqrt{s}=2.44$; (c)~off resonance, $g=0.8$, $\sqrt{s}=2.66$; (d)~free theory, $g=0$, $\sqrt{s}=2.42$. Upper row: excess charge $\delta\langle n_{j}\rangle$, positive (red) for particles entering from the left and negative (blue) for antiparticles entering from the right. Middle row: the site-resolved scalar count $n_{\rm inc}(j)$; the lump at the centre is the mediator. Lower row: the entanglement entropy across bond $j$ relative to the vacuum, $\Delta S_{j}$, in bits. In the upper two rows, adjacent sites are averaged to remove the two-site staggering. Dotted lines: the collision time $t_{c}=(x_{0,R}-x_{0,L})/(2v_{g})$.}
    \label{fig:ffbar-spacetime}
\end{figure*}

We now turn our attention to fermion--antifermion scattering, where the scalar appears as a resonance in the $s$ channel, $\psi\bar\psi\to\varphi\to\psi\bar\psi$. We arrange for a particle packet to enter from the left and an antiparticle packet from the right with central momenta $\pm k_{0}$, each moving at the group velocity $v_{g}=\mathrm{d}\omega/\mathrm{d}k$ at $k_{0}$, so the pair collides at zero total momentum and invariant mass $\sqrt{s}=2\omega(k_{0})$. The pair can annihilate into the scalar, which may then decay. We use $N=72$, $\sigma_{k}=0.1$ and packets centred at $x_{0}=20$ and $52$, and evolve to $t=140$. At the bare mass, the resonance sits at $k_{0}=\pi/4$. At one loop, its pole lies at $M=2.398$ with $\Gamma_{\rm D}=0.0996$ for $g=0.8$, and at $M=2.431$ with $\Gamma_{\rm D}=0.0407$ for $g=0.5$. 

In Fig.~\ref{fig:ffbar-spacetime} we present four emulated collisions. In panel (d), $g=0$, so the particle (red) and the antiparticle (blue) cross at the centre and continue unchanged. The entanglement entropy returns to its initial value once they have passed. Panel (a) is close to resonance. Here the outgoing stripes are faint, because each side of the vertex now holds a superposition of a transmitted and a reflected packet of opposite charge. A lump of scalar quanta forms at the vertex as the packets overlap and then fades, representing the mediator. Its count peaks at $t=51.6$, after the crossing at $t_{c}=38.8$. In panel (b), the scalar lump peaks later, at $t=56.8$, and survives to the end of the panel, since the mediator lives $2.4$ times longer than in panel (a). Panel (c) is off resonance; the stripes keep their colour, as most of the charge is transmitted, and only a faint trace of scalar quanta appears. At $t=100$, the entanglement entropy between the left and right halves of the chain, measured relative to the vacuum, is $1.63$, $1.96$ and $0.80$ bits for collisions (a), (b) and (c) respectively, compared with zero in the free theory.

\begin{figure}[t]
    \centering
    \includegraphics[width=0.95\linewidth]{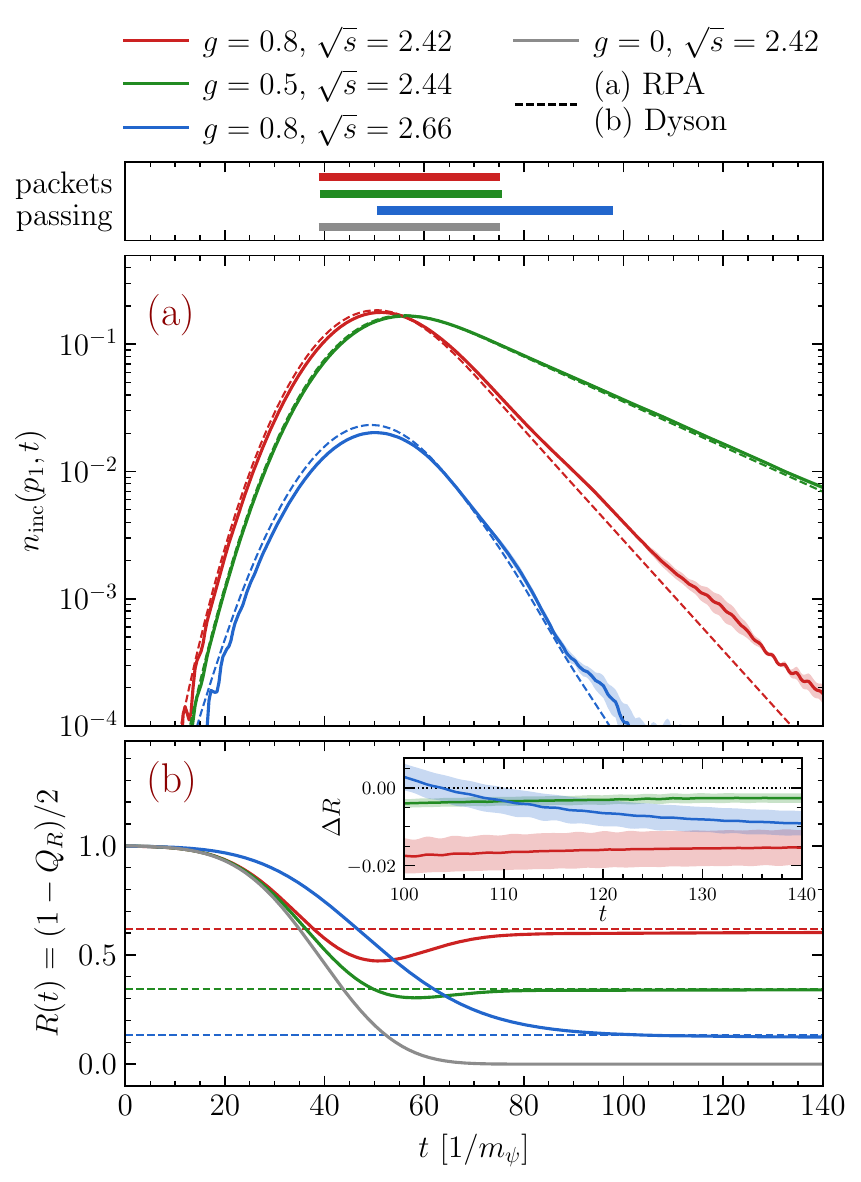}
    \caption{The collisions of Fig.~\ref{fig:ffbar-spacetime} at $\Lambda=6$ and to $t=140$. Top strip: the time during which the packets pass through each other, from $t_{c}$ to $t_{a}=t_{c}+3\sigma_{x}/v_{g}$, with $\sigma_{x}=1/(2\sigma_{k})$. (a)~The occupancy of the slowest motional mode, $n_{\rm inc}(p_{1},t)$ with $p_{1}=\pi/73$, computed in the MPS (solid) and RPA (dashed). (b)~$R(t)$, the charge-reflection observable. Inset: $\Delta R = R(t)-R_{\rm Dyson}$ at late times. Shaded bands: numerical uncertainty (cutoff, bond dimension and time step, added in quadrature).}
    \label{fig:ffbar-vertex}
\end{figure}

Figure~\ref{fig:ffbar-vertex} shows the time evolution of the same collisions. Solid lines are the MPS emulations, with on-resonance in red, the longer-lived mediator in green, off-resonance in blue, and the free theory in grey. The bars above panel~(a) mark the passage of the packets through each other, from $t_{c}=(x_{0, R}-x_{0, L})/2v_{g}$ to $t_{a}=t_{c}+3\sigma_{x}/v_{g}$. All packets start at the same positions, so the off-resonance bar starts later and lasts longer because its packets are slower, with $v_{g}=0.317$ against $0.412$ on resonance. In the continuum, they would be faster, but the lattice band flattens towards the edge of the Brillouin zone.

Panel~(a) shows the incoherent occupation of the slowest motional scalar mode, $n_{\rm inc}(p_1=\pi/73,t)$, which captures a mediator produced at rest. Dashed lines show the RPA evolution. The free theory value stays below $10^{-4}$, which is off the scale. On resonance (red), the mode fills as the packets overlap and peaks at $0.178$ quanta, close to the RPA value of $0.185$. At $g=0.5$ (green), the corresponding peak is $0.167$, compared with $0.169$ in the RPA. Off resonance (blue), the mediator remains virtual. Its mode occupation follows the overlap of the incoming packets and falls as they separate, with no subsequent decay. On resonance, by contrast, the mediator continues to decay after the packets have passed through one another, and at late times the red solid curve lies significantly above the corresponding RPA curve. We extract the decay rate again by fitting $A\,e^{-\Gamma t}+C$ over $t_a\le t\le140$ to both the emulation and the RPA evolution. At $g=0.8$, we obtain $\Gamma=0.092$, corresponding to $0.92$ times the RPA value. At $g=0.5$, we find $\Gamma=0.0398$, which is $0.97$ times the RPA value. The emulated mediator decays more slowly than the RPA would predict, which is consistent with the results we found when considering the scalar decay.

Panel (b) shows the charge-reflection observable $R(t)$.  The solid curves are the MPS results and the horizontal dashed lines of the same colour are the expectations based on explicit calculation of the reflection probability, $R(t\to\infty)$, using one-loop Dyson summation and including the tree-level $t$-channel scalar exchange contribution (labelled `Dyson'). $R(t)$ starts at unity since the fermion propagates in from the left. It then falls as the particle crosses to the right. In the free theory (grey), it falls monotonically to zero, since both packets pass through unchanged. On resonance (red), it first drops below its final value and then recovers. The parts of the packets that do not form the mediator pass through each other as in the free theory and lower $R(t)$. The rest is held by the neutral mediator and re-emitted as it decays; the emitted wave adds reflected charge and, in the forward direction, cancels part of the transmitted wave, so $R(t)$ rises again. The minimum of $R(t)$ coincides with the peak of the mediator in panel~(a), at $t\approx51$, and at $g=0.8$, $R(t)$ then approaches its final value at a rate close to $\Gamma$. At $g=0.5$ (green), the same dip appears at $t\approx58$, but the recovery is faster than $\Gamma$. For off-resonance (blue), there is no mediator, so there is no dip. The inset magnifies the difference $\Delta R = R(t)-R_{\rm Dyson}$ for $t\ge100$. 

\begin{figure}[t]
    \centering
    \includegraphics[width=0.95\linewidth]{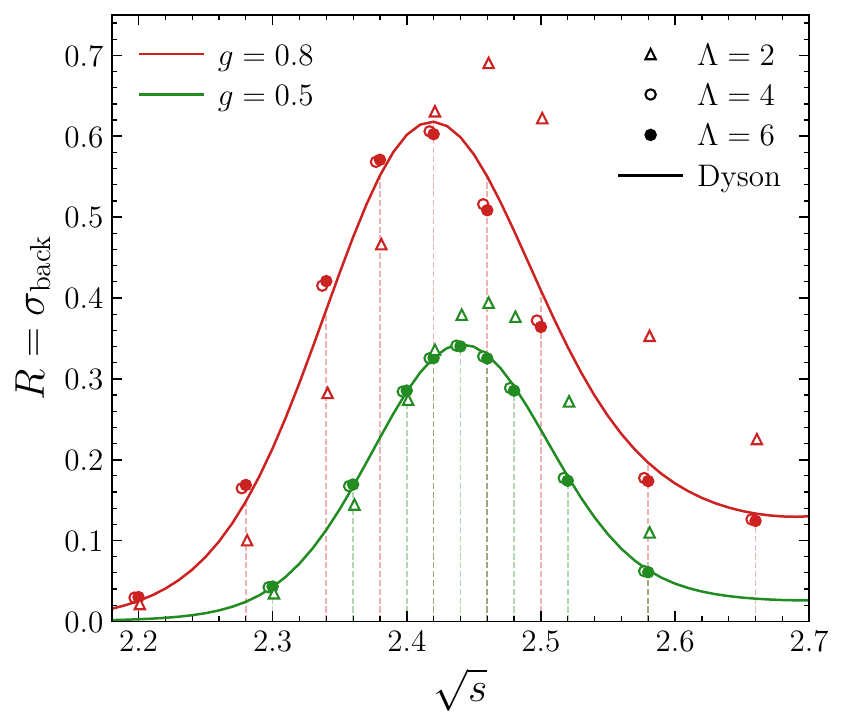}
    \caption{Resonance lineshape: the backward cross section $\sigma_{\rm back} = R(t=140)$ against the invariant mass $\sqrt{s}$ of the colliding pair for $g=0.8$ (red) and $g=0.5$ (green), with $\Lambda=2$ (open triangles), $\Lambda=4$ (open circles) and $\Lambda=6$ (filled circles) Fock levels per site ($N=72$, $\sigma_{k}=0.1$, $\chi=64$). The points are drawn slightly displaced in $\sqrt{s}$ for visibility (the vertical dashed lines mark the sampled energies). The solid curves correspond to the Dyson prediction averaged over the momentum spread of the packets.}
    \label{fig:ffbar-lineshape}
\end{figure}

Figure~\ref{fig:ffbar-lineshape} shows the collision-energy scan across the resonance. Each point is the reflection $R(t=140)$ determined at the end of a single run. Over the interval $119\le t\le140$, $R(t)$ changes by at most $0.003$, so the effect of the finite chain is minimal. The solid curves are the Dyson predictions, averaged over the momentum spread of the packets. The packets have an energy spread of $0.058$, comparable to $\Gamma$, and this lowers the peaks below the unitarity limit of 1 to $0.618$ at $g=0.8$ and $0.343$ at $g=0.5$. At $g=0.5$, the $\Lambda=6$ points agree with the curve within $0.006$ at every energy. At $g=0.8$, they are displaced to lower energy. 

Figure~\ref{fig:ffbar-lineshape} also shows how the Fock cutoff affects the resonance. With a single qubit per site, $\Lambda=2$, the resonance is misplaced and broadened, in line with the overestimated decay widths at $\Lambda=2$ in Table~\ref{tab:decay-width}. From $\Lambda=4$ on, the points change little, because the scalar field barely fills the site oscillators. In the $\Lambda=4$ and $6$ collisions of Fig.~\ref{fig:ffbar-lineshape}, the mean occupation $\langle a_{j}^{\dagger}a_{j}\rangle$ of a site oscillator, over all sites, times and energies, peaks at $0.061$ for $g=0.8$ and $0.051$ for $g=0.5$, both at the vertex of the collisions near the resonance. In small-scale exact-diagonalisation tests, the levels $n\le3$ retained at $\Lambda=4$ hold all but about $3\times10^{-5}$ of the probability at every site.

The small change from $\Lambda=4$ to $6$ is nevertheless systematic, and the whole lineshape moves slightly down in energy. Its origin is the tail of the occupation. The motional vacuum is squeezed and displaced relative to the site oscillators, so a site already holds one or two quanta with a small probability. When the mediator forms, its quantum adds to these, populating $n=3$, the highest level kept at $\Lambda=4$. The Rayleigh--Ritz projection keeps the energy of this level exact but removes its couplings to the levels above, and such couplings can only lower the energy~\cite{MacDonald_1933}. The truncation therefore raises the mediator, but not the fermions it decays into, and the resonance moves up. Exact-diagonalisation and MPS analyses confirm that the shift is carried by the scalar and disappears once the $n=4$ level is added. At $g=0.5$, this upward shift almost cancels the small downward displacement of the converged resonance from Dyson, so the closer agreement of $\Lambda=4$ with Dyson is accidental. Without the Rayleigh--Ritz projection, squaring the truncated $\varphi_{j}$ and $\Pi_{j}$ instead, the emulation produces $R(t=140)\approx0$ at the cutoffs we tested, $\Lambda\in\{2,3,4\}$.

%%%%%%%%%%%%%%%%%%%%%%%%%%%%%%%%
\section{Conclusion}
\label{sec:conclusion}
%%%%%%%%%%%%%%%%%%%%%%%%%%%%%%%%

In this study, we formulated real-time scattering in the $(1+1)$-dimensional Yukawa theory, the simplest field theory in which fermions couple to a scalar, for a hybrid qubit--qumode device, and emulated it with MPS. On trapped ions, the fermions are held by the internal states of the ions and the scalar by the collective motional modes of the chain. In this encoding, the Yukawa coupling is a spin-dependent displacement, a native trapped-ion operation, and every observable that we compare with theory is a native measurement. With this set-up, we followed two processes in real time, the decay of a single scalar quantum into a fermion--antifermion pair and the resonant scattering of a fermion--antifermion pair through the scalar.

Our emulation is precise enough to resolve departures from one-loop lattice perturbation theory with Dyson resummation. The scalar, whether prepared directly or produced by the colliding pair, decays $2$--$8\%$ more slowly than in the Dyson-resummed one-loop evolution, consistent with the next-to-leading-order width. The reflected charge reproduces the resummed lineshape at \mbox{$g=0.5$}, whereas at $g=0.8$ the resonance lies slightly lower in energy than predicted. The same simulations show how many Fock levels the scalar needs. With two levels per site, the content of one qubit, the decay width is overestimated by up to $30\%$, in the opposite direction to the next-to-leading-order correction, and the resonance is misplaced. Four levels resolve the width correction, but the position of the resonance is stable only from six levels on, because a smaller cutoff removes part of the self-energy of the mediator and raises its mass.

Our results show the potential importance of hybrid qubit--qumode systems for simulating theories with fermionic and bosonic fields. Even at these weak couplings, the scalar needs more levels than a single qubit, and we expect the number to grow with the collision energy and the coupling. A binary encoding of such a scalar in the harmonic-oscillator basis would require $\mathcal{O}(\log^2\Lambda)$ entangling gates per interaction term~\cite{DavoudiLinkePagano2021}. On a trapped-ion device, each Trotter layer is a native operation, and a similar structure carries over to circuit-QED, where an echoed conditional displacement plays the role of the spin-dependent force~\cite{Blais2021, Eickbusch2022}. The collisions shown here would need $1400$ Trotter steps on $72$ idealised ions, assuming $\delta t=0.1$, with about $2\times10^{5}$ M\o{}lmer--S\o{}rensen gates and, in the normal-mode frame, about $7\times10^{6}$ conditional displacements, so running them will require error rates well below those achieved today. 

The collisions studied here remain within reach of a classical emulation because the couplings are weak, the boson occupancies are low, and the times are moderate. The natural next step is studying the non-perturbative regime. At stronger coupling, the resonance broadens and mixes with the two-fermion continuum, and dressing the fermions and their interactions is no longer a small correction, so perturbative benchmarks like those used here lose reliability. Higher collision energies open inelastic channels, increase boson occupancies, and generate more entanglement~\cite{Rigobello:2021fxw, Belyansky:2023rgh}, and the cost of an MPS emulation grows exponentially with entanglement. For such collisions followed for long times, or in more than one spatial dimension, a quantum device becomes essential. Encoding fermions on qubits and bosonic fields on qumodes carries over to other theories in which fermions couple to bosonic fields, including gauge theories.

\paragraph*{Acknowledgements.} We used Anthropic's Claude Opus 5.5 and OpenAI's GPT-6.1 Sol in this project to search the parameter space, parallelise the benchmarks on the cluster, and check calculations. The authors checked all results and take full responsibility for the content. The numeric simulations use NumPy~v2.4.6~\cite{harris2020array} and SciPy~v1.17.1~\cite{2020SciPy-NMeth} with the MKL implementation of the BLAS library. JYA is supported by the Institute for Particle Physics Phenomenology Associateship Scheme and by the Royal Society under grant no. IES/R2/252139. RJ would like to thank Callsign Ltd for their financial support. 

\paragraph*{Note added.}
While this work was being completed, Ref.~\cite{Abel:2026gao} appeared, proposing a photonic qubit--qumode implementation of Yukawa theory and also studying resonant fermion--antifermion scattering with tensor networks.

\bibliography{bibliography}
\end{document}